\documentclass[superscriptaddress,twocolumn,aps,prd,10pt,nofootinbib,preprintnumbers]{revtex4-2}

\usepackage{graphicx}
\usepackage{amsmath} 
\usepackage{amssymb}
\usepackage[colorlinks=true, pdfstartview=FitV, linkcolor=red, citecolor=blue, urlcolor=blue, bookmarks=false]{hyperref}
\usepackage{here}

\newcommand{\rmd}{\mathrm{d}}
\newcommand{\bx}{\boldsymbol{x}}

\begin{document}

\title{Relativistic BEC extracted from a complex FRG flow equation}

\author{Fumio~Terazaki}
\email{1221709@ed.tus.ac.jp}
\affiliation{Department of Physics, Tokyo University of Science, Tokyo 162-8601, Japan}
\author{Kazuya~Mameda}
\affiliation{Department of Physics, Tokyo University of Science, Tokyo 162-8601, Japan}
\affiliation{RIKEN iTHEMS, RIKEN, Wako 351-0198, Japan}
\author{Katsuhiko~Suzuki}
\affiliation{Department of Physics, Tokyo University of Science, Tokyo 162-8601, Japan}

\preprint{RIKEN-iTHEMS-Report-24}

\begin{abstract}%
Based on the functional renormalization group (FRG) under the local potential approximation, we analyze the Bose-Einstein condensation (BEC) in the relativistic complex scalar theory.
This framework leads to a complex flow equation of the effective potential, even with the well-known Litim regulator.
In order to evaluate the condensate from such a complex effective potential, we impose a condition between chemical potential and mass, analogously to those in the free theory or the mean field theory.
We elucidate that for the strongly (weakly) coupled theory, the phase diagrams computed from the FRG are more (less) deviated from that under the mean field approximation.
This result implies that quantum fluctuations strongly affect the nonperturbative formation of the BEC.
\end{abstract}

\maketitle

\section{Introduction}
The Bose-Einstein condensation (BEC) of relativistic field theories is as intriguing as its occurrence in nonrelativistic systems, such as cold atomic gases~\cite{Dalfovo:1999zz,pethick2008bose}.
The relativistic pion and kaon condensates~\cite{Sawyer:1972cq,Barshay:1973ju,Baym:1973zk,Kaplan:1986yq} have the potential to influence the structure of neutron stars, though it is still under debate whether these BECs exist in the interior~\cite{Pethick:2015jma,Malik:2020jlb}.
Behaviors of the latter condensate have been analyzed in the color-flavor locking phase~\cite{Alford:1998mk}, which would be realized in the center of neutron stars.
There are many studies about the BEC as a candidate for dark matter~\cite{Urena-Lopez:2008vpl}; this is also argued in the context of the nonrelativistic BEC~\cite{Boehmer:2007um}.

There are several approaches to analyzing the relativistic BEC of interacting complex scalar fields.
One is based on the mean field approximation (MFA).
As in nonrelativistic BECs, this approach provides the self-consistent equation of the condensate, and yields the analytical formula of the critical temperature dependent on the chemical potential~\cite{Kapusta:1981aa}.
The one-loop approximation is a different approach from the MFA~\cite{Bernstein:1990kf,Sharma:2022jio,Nicolis:2023pye}, but leads to the same critical temperature~\cite{Benson:1991nj}.
Unfortunately, these tractable approaches for the relativistic BEC do not capture the quantum fluctuation, which potentially becomes crucial around the second-order phase transition.

Although the functional renormalization group (FRG)~\cite{Wetterich:1992yh,Berges:2000ew,Bagnuls:2000ae,Wetterich:2001kra,Gies:2006wv,Kopietz:2010zz,Dupuis:2020fhh} is a powerful tool to examine the significance of quantum fluctuations, its practical usage sometimes brings nontrivial issues.
The most relevant one to the relativistic scalar theories is the emergence of the complexity of the effective potential.
For instance, in the FRG of the quark-meson model, an imaginary part of the flow equation (and thus of the resulting effective potential) emerges in some region of the coupling parameter~\cite{Schaefer:2004en}.
The same is true for the finite-density complex scalar theory~\cite{Svanes:2010we}.
In the above cases, one can keep the effective potential real, by tuning the coupling parameters or just ignoring its imaginary part.
It is, however, unclear how these prescriptions are justified.

An alternative scheme to extract the relativistic BEC from a complex flow equation is to impose a condition between chemical potential and an effective mass.
This is more plausible physically in the sense that a similar constraint is inherently required in the free scalar theory or the MFA~\cite{Kapusta:1981aa,Kapusta:2006pm}.
In this paper, we analyze the relativistic BEC with the above scheme, and investigate the effect of the quantum fluctuations.
We compute the phase diagram of the interacting complex scalar field theory based on the FRG under the local potential approximation.
Although the complex scalar theory in four-dimension is quantum trivial at vacuum~\cite{Frohlich:1982tw,Aizenman:2019yuo,Romatschke:2023sce}, it is subtle whether this is also true at finite temperature and density.
We indeed observe that the strong coupling case is largely deviated from the MFA due to quantum fluctuations, while the weak coupling case converges to the MFA result.

This paper is organized, as follows. 
In Sec.~\ref{sec:formalism}, we derive the flow equation, and explain the method to extract the ground state from the complex flow equation.
In Sec.~\ref{sec:result}, we show the flow of the potential evaluated with the above method.
Also we compare the phase diagrams with the MFA and the FRG, and investigate the effects of quantum fluctuations.
In Sec.~\ref{sec:summary}, we summarize our findings, and mention outlooks.
Appendix~\ref{sec:app} is devoted to the detailed derivation of the flow equation of the effective potential.


\section{FRG flow equation}\label{sec:formalism}
In the imaginary-time formalism, the complex scalar field theory is described by the following action~\cite{Kapusta:1981aa}:
\begin{equation}
\begin{split}
S
&= \int_x \varphi^* 
\Bigl[
-(\partial_{\tau}-\mu)^2-\nabla^2+\bar{m}^2 
+\bar{\lambda} |\varphi|^2
\Bigr]\varphi,
\end{split}
\label{eq:S}
\end{equation}
with $\int_x = \int_{0}^{\beta}\rmd\tau\int\rmd^3\bx$ and $x^\mu = (\tau,\bx)$.
Here $\varphi=\varphi(x)$ is a complex scalar field with a mass $\bar{m}$ and the coupling is $\bar{\lambda}$.
Also we introduce temperature as $\beta^{-1}=T$, and
$\mu$ is the chemical potential for the $U(1)$ charge density $j^0$, where $j^\mu = \mathrm{i} (\varphi^* \partial^\mu \varphi - \varphi\partial^\mu \varphi^*)$.
In this paper, we restrict the positive chemical potential $\mu\geq 0$, as the theory is invariant under the charge-conjugation.

The FRG is a method to incorporate quantum fluctuations nonperturbatively.
We define $\Gamma_k$ as the effective action obtained by integrating out higher modes up to a momentum scale $k$.
The classical action $S=\Gamma_{k=\Lambda}$ (with the ultraviolet scale $\Lambda$) and the quantum effective action $\Gamma_{k=0}$ are interpolated by $\Gamma_k$, which obeys the flow equation~\cite{Wetterich:1992yh,Berges:2000ew,Bagnuls:2000ae,Wetterich:2001kra,Gies:2006wv,Kopietz:2010zz,Dupuis:2020fhh}:
\begin{equation}
\frac{\partial \Gamma_{k}}{\partial k}
=\frac{1}{2}{\rm Tr}\left[\frac{\partial R_{k}}{\partial k}\frac{1}{\Gamma_{k}^{(2)}+R_{k}}\right].
\label{eq:Wfloweq}
\end{equation}
The two-point vertex function $\Gamma_{k}^{(2)}$ is a matrix defined as
\begin{equation}
[\Gamma_{k}^{(2)}(p,q)]_{ij}= \frac{\delta^2\Gamma_{k}}{\delta\Phi_{i}^{\dagger}(p)\delta\Phi_{j}(q)},
\quad
\Phi(p) =
\begin{pmatrix}
\varphi(p) \\ \varphi^*(-p)
\end{pmatrix}.
\end{equation}
We denote the momentum valuable as $q = (\omega_{n}, \boldsymbol{q})$, where the bosonic Matsubara frequency $\omega_{n}=2\pi n T$.
Also $\rm{Tr}$ is the trace with respect to the momentum indices $(p,q)$ and the field indices $i,j$.
In this paper, we employ the Litim regulator~\cite{Litim:2001up}:
\begin{equation}
R_{k}(p,q)=\delta^{(4)}(p-q)\,(k^2-\boldsymbol{p}^2)\,\theta(k-|\boldsymbol{p}|),
\label{eq:Litim}
\end{equation}
where we define $\delta^{(4)}(p-q):=(2\pi)^3\beta\delta_{n_{p},n_{q}}\delta^{(3)}(\boldsymbol{p}-\boldsymbol{q})$.

The flow equation~\eqref{eq:Wfloweq} cannot be evaluated in a closed form;
the flow of $n$-point functions $\Gamma_{k}^{(n)}$ is obtained from an equation involving $\Gamma_{k}^{(n+1)}$ and $\Gamma_{k}^{(n+2)}$~\cite{Berges:2000ew}.
To truncate the flow equation, we here perform the derivative expansion~\cite{Golner:1985fg}, together with applying the local potential approximation~\cite{Nicoll:1974zz}.
Then, we write the effective action as
\begin{equation}
\begin{split}
\Gamma_{k}
&= \int_x\biggl[\varphi^*\Bigl(-(\partial_{\tau}-\mu)^2-\nabla^2\Bigr)\varphi
+ V_k(|\varphi|^2)
\biggr].
\end{split}
\label{eq:Gammak}
\end{equation}
Here $V_{k}$ includes any nonnegative order terms of $|\varphi|^2$, so that the $U(1)$ global symmetry is respected as in Eq.~\eqref{eq:S}.
Furthermore, let us suppose that ground states are translation-invariant, and represented by a homogeneous field $\phi$.
From the flow equation~\eqref{eq:Wfloweq} of $\Gamma_k$, then, we can extract that for the following effective potential:
\begin{equation}
U_{k}(\rho)
:=V_{k}(\rho)-\mu^{2}\rho,
\quad
\rho := |\phi|^2 .
\label{eq:Uk}
\end{equation}
The resulting form is
\begin{equation}
\frac{\partial U_{k}}{\partial k}
=\frac{k^4}{6\pi^2}
\sum_{s=\pm}
\frac{E_{s}^2+\mu^2-E_{k}^2}{E_{s}^2-E_{-s}^2}\frac{1+2n_\mathrm{B}(E_{s})}{E_{s}},
\label{eq:floweq}
\end{equation}
where $n_\mathrm{B}(x)=(\mathrm{e}^{\beta x}-1)^{-1}$ and
\begin{equation}
\begin{split}
&\qquad E_{k}^2=k^2+\mu^2+U_{k}'+\rho U_{k}'',\\
&E_{\pm}=\sqrt{E_{k}^2+\mu^2\pm\sqrt{4E_{k}^2\mu^2+(\rho U_{k}'')^2}}.
\label{eq:Es}
\end{split}
\end{equation}
The prime symbols on $U_k$ denote the derivatives with respect to $\rho$.
The derivation is shown in Appendix~\ref{sec:app}.
The same flow equation is derived in Ref.~\cite{Palhares:2012fv}, although the notations are different.
The flow equation~\eqref{eq:floweq} is numerically solved under the initial condition $\Gamma_{k=\Lambda}=S$, which implies
\begin{equation}
U_{k=\Lambda}(\rho)=\bar{m}^2\rho+\bar{\lambda}\rho^2-\mu^2\rho.
\label{eq:ULambda}
\end{equation}
The momentum scale of the theory is characterized by the bare mass $\bar{m}^2=V'_{k=\Lambda}(\rho=0)$.
All dimensionful parameters $\bar{m}$, $T$ and $\mu$ are taken to be much smaller than the ultraviolet scale $\Lambda$.

The ground state is identified through 
\begin{equation}
\rho_{0}:=\underset{\rho}{\mathrm{arg\,min}} \,U_{k\to 0}(\rho).
\label{eq:rho0}
\end{equation}
This is the order parameter of the BEC phase transition;
$\rho_0 >0$ corresponds to the BEC phase, while $\rho_0 =0$ is the normal phase.
From this criterion, we can draw the phase diagram on the $(T,\mu)$-plane.
Then, a good reference for the FRG analysis is the phase transition lines under the MFA~\cite{Kapusta:1981aa}:
\begin{equation}
T_{\rm c}^2=\frac{3}{\bar{\lambda}}\left(\mu_{\rm c}^2-\bar{m}^2\right)
\quad({\rm MFA})
\label{eq:Tc}
\end{equation}
with $T_{\rm c}$ and $\mu_{\rm c}$ being the critical temperature and chemical potential, respectively.
The formula~\eqref{eq:Tc} is unchanged by adding several missing one-loop contributions~\cite{Benson:1991nj}, but in the following we term it the MFA result.
An important remark is that the mass and coupling parameters involved here are those of the classical action, that is,  $\bar{m}$ and $\bar{\lambda}$, respectively.
Hence, Eq.~\eqref{eq:Tc} is directly compared with the phase diagram computed by the flow equation~\eqref{eq:floweq} with the same $\bar{m}$ and $\bar{\lambda}$;
see also Ref.~\cite{Fukushima:2012xw}, where a similar comparison is made for the magnetic property of the chiral condensate.

A remarkable feature in the flow equation~\eqref{eq:floweq} is the complexity.
For instance, it is readily found that $E_-$ becomes imaginary for a large $\mu$.
We emphasize, however, that this feature is not specific to the FRG framework.
In fact, in relativistic scalar theories, there is a constraint on a mass and chemical potential: $\mu^2\leq ({\rm  mass})^2$, where $(\mathrm{mass})$ is the bare mass in the free case and an effective mass under the MFA~\cite{Kapusta:1981aa,Kapusta:2006pm}.
This bound comes from the requirement that the thermodynamic potential be real and finite, or that the Bose distribution be positive.
A similar requirement should be imposed in the thermodynamics derived from the FRG framework.
Therefore, the effective potential $U_k(\rho)$, and thus the right hand side of Eq.~\eqref{eq:floweq} should be real and finite.
For $k>0$, this is respected as long as $E_-^2>0$.
We note that $E_- = 0$ is not allowed by the finiteness of the right hand side in Eq.~\eqref{eq:floweq}.
This inequality holds if%
~\footnote{%
The inequality $E_-^2>0$ is also fulfilled when $\mu^2> k^2+V_{k}'+2\rho V_{k}''$.
Nevertheless, we have numerically confirmed that this condition never holds for our parameter choices as $\bar{m},T,\mu \ll \Lambda$.
}
\begin{equation}
\mu^2 < k^2+V_{k}'=:m_{\mathrm{eff},k}^2,
\quad 
k > 0.
\label{eq:condition}
\end{equation}
It is obvious that the condition~\eqref{eq:condition} is the FRG counterpart of the bound in the free theory or the MFA.
For $k\to 0$, it might be expected that Eq.~\eqref{eq:condition} is reduced to $\mu^2 < V_{k\to 0}'$.
However, one can analytically show that as long as $U_k'$ goes to zero faster than $k$ does, the flow equation is still regular even for $E_- = 0$.
Hence, the above condition is slightly relaxed to
\begin{equation}
 \mu^2 \leq V_{k\to 0}',
\quad 
k \to 0.
\label{eq:condition_IR}
\end{equation}
In the following numerical analysis, we exclude the $\rho$ that does not satisfy Eq.~\eqref{eq:condition} or~\eqref{eq:condition_IR}.

The necessity of the above conditions is one of the main findings in this paper, unlike the preceding relevant works.
For instance, in Refs.~\cite{Svanes:2010we} and~\cite{Palhares:2012fv}, their flow equations are basically the same as ours, but the above conditions are not taken into account.
In the former, the imaginary part in the flow equation is dropped by hand.
In the latter, the author chooses the parameters of the bare mass and coupling with which the imaginary part disappears.
We also emphasize that since $\mu^2\leq (\mathrm{mass})^2$ is required in order that the thermal Bose distribution is well-defined, the complexity in the FRG flow equation would be specific to the finite-temperature bosonic theory.
Indeed, if the momentum phase space were $O(4)$-symmetric unlike ours over $q=(\omega_n,\boldsymbol{q})$, the flow equation would be always real~\cite{Berges:2000ew}.

The above conditions indicate the limitation of the Litim regulator.
In general, the effective potential becomes ill-defined when its slope and curvature of the effective potential is nonpositive for $k\neq 0$ or negative for $k\to 0$~\cite{Litim:2000ci}.
One of the crucial properties of the Litim regulator is to evade this pathological situation, as much as possible~\cite{Litim:2001up}.
In the relativistic complex scalar theory at finite density, however, such a well-defined effective potential is not guaranteed only by employing the Litim regulator~\eqref{eq:Litim}, as seen in Eq.~\eqref{eq:floweq}.
Then, we need to explicitly impose the conditions of the positivity $U'_k+k^2>0$, and of the nonnegativity $U'_{k\to 0}\geq 0$, which are equivalent to the conditions~\eqref{eq:condition} and~\eqref{eq:condition_IR}, respectively.

Finally, we briefly mention the numerical method to solve Eq.~\eqref{eq:floweq}.
First, it is useful to introduce the dimensionless valuables $t=\log(k/\Lambda)$ and $r=\rho/\Lambda^2$.
We then employ the fourth-order Runge-Kutta method for the $t$-derivatives from $t=0$ to $t=\log(10^{-4})$%
~\footnote{
The corresponding value $k=10^{-4}\Lambda$ is much smaller than $k\simeq \bar{m}$, around which our numerical computation of the FRG flow converges well.
This choice is only for the sake of drawing Figs.~\ref{fig:potential_flow} and~\ref{fig:rholeft}.
}, and the grid method~\cite{Adams:1995cv,Schaefer:2004en,Drews:2016wpi} for the $r$-derivatives.
The discretization widths are chosen as $\Delta t\simeq-4.6\times10^{-7}$ and $\Delta r = 10^{-4}$, and the number of the $t$-integration steps is $2.0\times10^7$.
In order to accurately locate $\rho_{0}$, we also interpolate adjacent grids with the cubic-spline.

Also we have to set a range of $\rho$ for numerical calculations, and the right endpoint $\rho_\mathrm{right}$ and the left endpoint $\rho_\mathrm{left}$.
This $\rho_\mathrm{left}$ is zero or determined from Eq.~\eqref{eq:condition} or~\eqref{eq:condition_IR}, while $\rho_\mathrm{right}$ can be arbitrary as far as $\rho_\mathrm{right}\gg \bar{m}^2,\mu^2,T^2$.
The derivatives at $\rho=\rho_\mathrm{left}$ ($\rho=\rho_\mathrm{right}$) endpoint of the $r$-grids are calculated with the forward (backward) differences, but the others with the central difference.
In the present analysis such a treatment becomes crucial, since the left endpoint gives the ground state, i.e., $\rho_0=\rho_\mathrm{left}$;
see the argument in the following section.
We note that for $E_-=0$, there emerges a numerical instability because the flow equation~\eqref{eq:floweq} diverges.
At a finite $k$, we approximately determine $\rho_\mathrm{left}$ as the $\rho$ leading to $|E_-|/\bar{m}\simeq 10^{-6}$.
However, as mentioned above Eq.~\eqref{eq:condition_IR}, $E_-=0$ for $U_{k\to 0}' = 0$ yields no numerical instability.
Therefore, such a numerical issue of $\rho_\mathrm{left}$ never affects the global minimum, which is determined from $U_{k\to 0}' = 0$.


\section{Numerical results}
\label{sec:result}

\begin{figure}
\centering
\includegraphics[width=0.8\columnwidth]{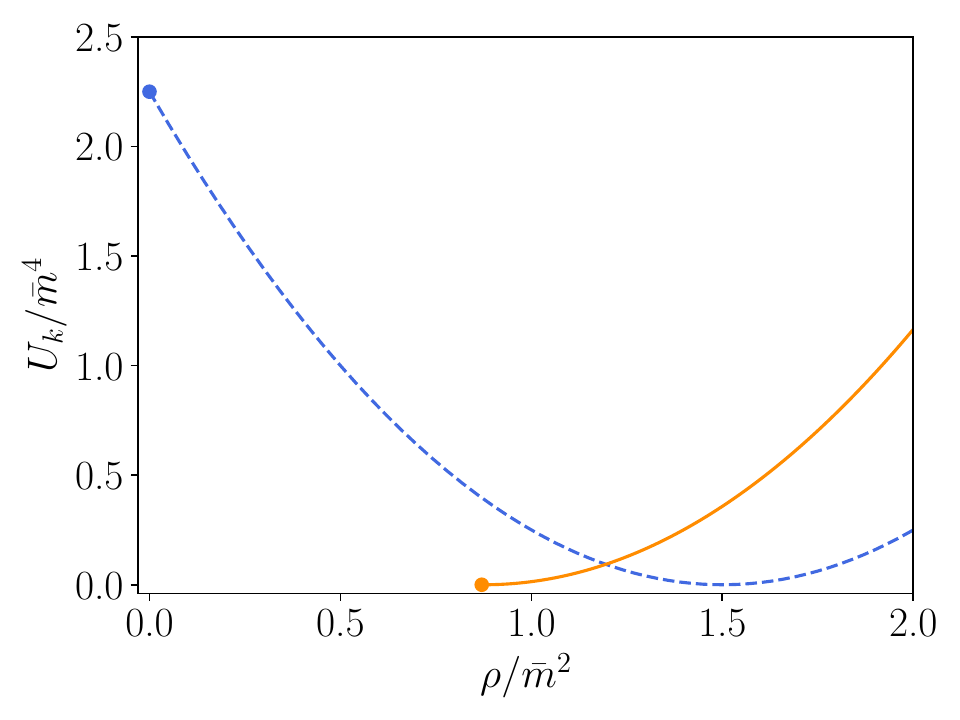}
\caption{
 Potential flow from the ultraviolet scale $k=\Lambda$ (dashed) to the infrared $k\to 0$ (solid).
}
\label{fig:potential_flow}
\end{figure}

\begin{figure}
\centering
\includegraphics[width=0.8\columnwidth]{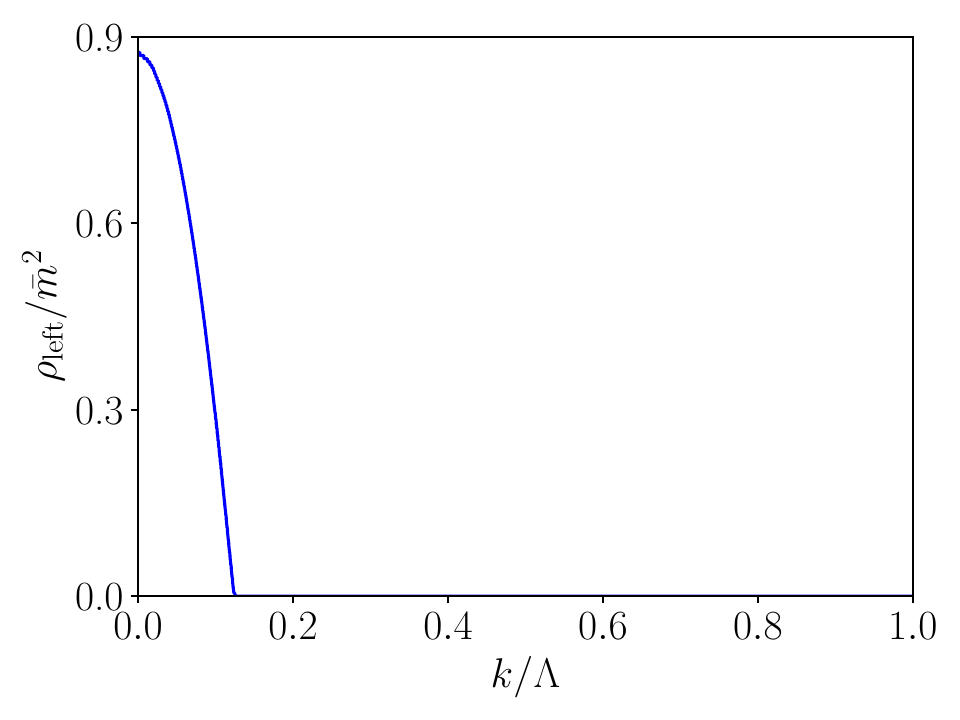}
\caption{
 Left endpoint of $\rho$ satisfying the condition~\eqref{eq:condition}.
}
\label{fig:rholeft}
\end{figure}

We now show the numerical result.
In the following, we fix the typical energy scale by
\begin{equation}
 \bar{m} = 0.1 \Lambda .
\end{equation}
In Fig.~\ref{fig:potential_flow}, we make a plot of the potential flow within the range of $\rho$ which fulfills Eq.~\eqref{eq:condition}.
We choose $T=0$, $\mu/\bar{m}=2.0$ and $\bar{\lambda}=1.0$, and the baseline is shifted by the minimum of $U_k$.
The dashed blue line is $U_{k=\Lambda}$ in Eq.~\eqref{eq:ULambda}, and the solid orange line is $U_{k\to 0}$, i.e., the solution of Eq.~\eqref{eq:floweq}.
Besides, the shift of the potential minima indicates that the order parameter is affected by quantum fluctuations.
The dots denote the left endpoint $\rho_{\rm left}$.
This endpoint is finite for the infrared $k$, while it is zero for the ultraviolet $k$.

In Fig.~\ref{fig:rholeft}, we show the $k$-dependence of the endpoint with the same parameters as in Fig.~\ref{fig:potential_flow}.
A crucial finding is that the left endpoint $\rho_{\rm left}$ is monotonically increased from the ultraviolet $k$ to the infrared $k$;
the onset is around $k\simeq \bar{m} = 0.1\Lambda$, as it should physically.
We have numerically checked that such a behavior is irrelevant to the parameter choices.
Once a grid point of $\rho$ is excluded by the condition~\eqref{eq:condition} at some $k$, e.g. $k = k_0$, we keep this point excluded for $k < k_0$.
This is a great advantage in the numerical computation of the phase diagrams.

\begin{figure}
\centering
\includegraphics[width=0.8\columnwidth]{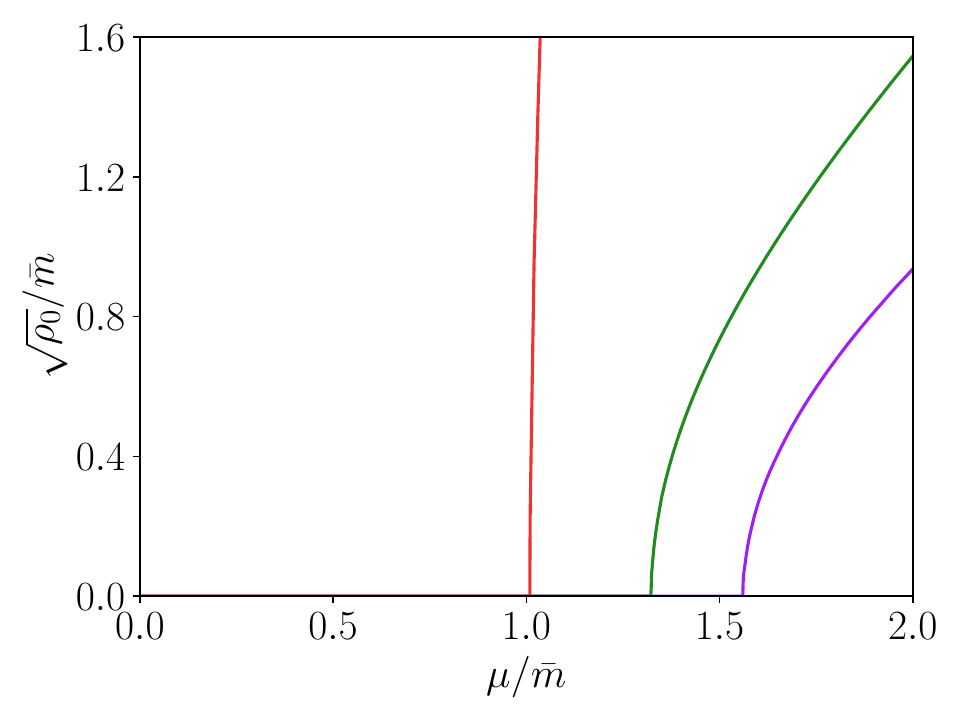}
\caption{
 Chemical potential dependence of the condensate at $T=0$.  We employ $\bar{\lambda} = 1.0$ (purple), $\bar{\lambda}=0.5$ (green) and $\bar{\lambda}=0.01$ (red).
}
\label{fig:rho0}
\end{figure}

In Fig.~\ref{fig:rho0}, we plot the order parameter at zero temperature as a function of the chemical potential.
The purple, green and red lines correspond to $\bar{\lambda}=1.0$, $\bar{\lambda}=0.5$ and $\bar{\lambda}=0.01$, respectively.
As the order parameter $\rho_{0}$ increases continuously, we confirm that the present FRG analysis leads to the second-order phase transition, similarly in Ref.~\cite{Svanes:2010we}.
Such a behavior is consistent with the analytical argument under the MFA~\cite{Kapusta:1981aa}.

\begin{figure}
\centering
\includegraphics[width=0.8\columnwidth]{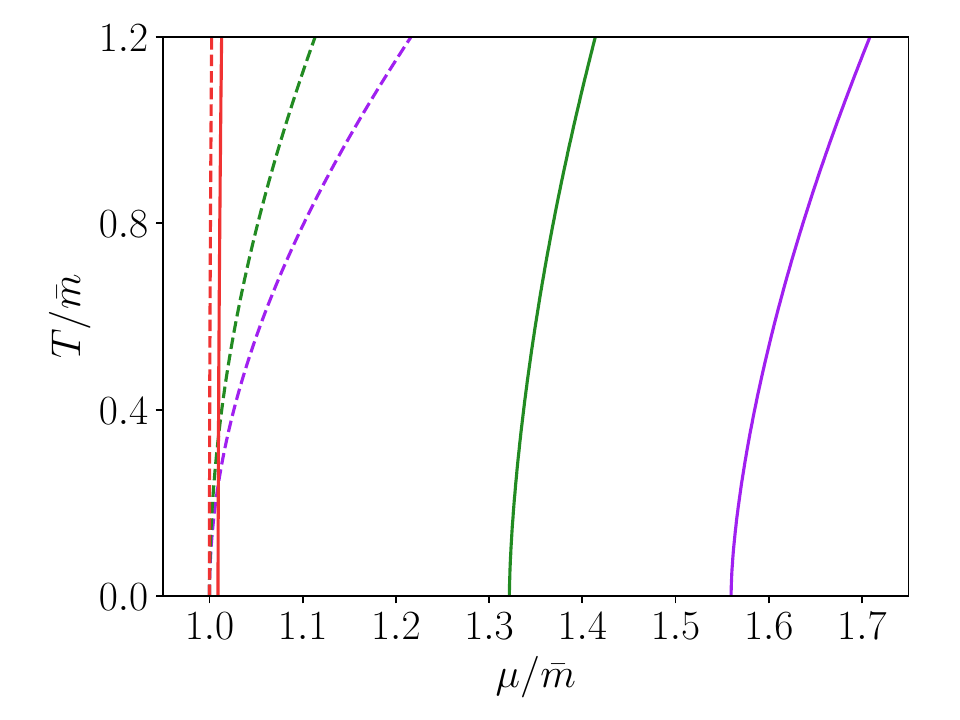}
\caption{
 Phase diagram with $\bar{\lambda}=1.0$ (purple), $0.5$ (green) and $0.01$ (red).
 The three solid (dashed) lines correspond to the FRG (MFA) results.
}
\label{fig:PTlines_real}
\end{figure}

Figure~\ref{fig:PTlines_real} shows the phase diagrams of the BEC transition.
While the solid lines represent the phase diagrams computed from the flow equation~\eqref{eq:floweq}, the dashed lines represent the MFA phase diagram described by Eq.~\eqref{eq:Tc}.
The correspondence between the color and $\bar{\lambda}$ is the same as Fig.~\ref{fig:rho0};
the three dashed lines intersect at $T=0$ because of $\mu_{\rm c}=\bar{m}$ in Eq.~\eqref{eq:Tc}.
Each right (left) region of the transition lines is the BEC phase with $\rho_{0}>0$ (the normal phase with $\rho_{0}=0$).
For a small $\bar{\lambda}$, the MFA and the FRG phase transition lines become closer.
Particularly, in the weak coupling limit $\bar{\lambda}\to 0$ (red lines),
both of the transition lines tend to converge to $\mu = \bar{m}$, which is that of the free theory~\cite{Kapusta:1981aa}.
On the other hand, the difference between the MFA and the FRG is enhanced for a large $\bar{\lambda}$.

The above result shows that the quantum fluctuation becomes crucial for the strong coupling, while it does not for the weak.
This is more nontrivial than it looks.
The MFA is in general unreliable not only in the strong coupling, but also at the vicinity of the critical point of the second-order phase transition.
For instance, the BCS mean field formula does not describe the correct behavior even in the weakly coupling limit due to the quantum fluctuation~\cite{Eberlein:2013nxa}.
Therefore, our computation numerically verifies the applicability of the MFA result~\eqref{eq:Tc}, at least under the local potential approximation.


\section{Summary}
\label{sec:summary}
We have investigated the relativistic BEC, using the FRG of the complex scalar theory.
In order to extract the real order parameter from a complex flow equation, we have imposed the condition between the effective mass and chemical potential.
This condition is the counterpart of the free theory or the MFA.
When the interaction is strong (weak), the phase diagram computed from the FRG
 is more (less) deviated from that under the MFA.
Hence, we have numerically demonstrated that in a strongly coupled regime, the MFA is inapplicable to the relativistic BEC, due to nonnegligible quantum fluctuations.
These results have, however, been obtained under the local potential approximation.
As the next step, we need to solve the coupled flow equations of the effective potential and the wave-function renormalization constant.
This approach could more quantitatively reveal the effect of quantum fluctuations on the relativistic BEC.

The present analysis is applicable to several directions.
One is the direct application to the relativistic high-density systems, i.e., neutron stars, where the lattice QCD simulation suffers from the sign problem.
A similar FRG analysis of the relativistic BEC would offer new insights into their inner structures including the color-flavor locking phase~\cite{Schafer:2001bq,Buballa:2004sx,Forbes:2004ww,Andersen:2008tn}.
Another is about the method developed in this paper;
we have determined the global minimum of the effective potential, by restricting the region where the potential is real.
The same method is helpful for other theories and models where the Litim regulator does not exclude an imaginary part of the effective potential.
As an advantage, this liberalizes the constraints on parameters used in the FRG analyses, and thus enables us to access the parameter region that has not been explored before.

\appendix

\section{Derivation of the flow equation}
\label{sec:app}
In this section, we show the critical steps to derive the flow equation~\eqref{eq:floweq}.
In order to compute the trace in the Eq.~\eqref{eq:Wfloweq}, we represent $\Gamma_{k}$ with the Fourier modes $\varphi(p)$ and $\varphi^{*}(p)$, which are defined as
\begin{equation}
\varphi(x)
=\int_{p}
{\rm e}^{{\rm i}\omega_{n}\tau+{\rm i}\boldsymbol{p}\cdot\boldsymbol{x}}\varphi(p),
\quad
\int_{p}:=T\sum_{n}\int\frac{\rmd^3p}{(2\pi)^3}.
\end{equation}
In the momentum space, the kinetic terms in the effective action are represented as
\begin{equation}
\begin{split}
&\int_x \varphi^*(x) 
\Bigl[
-(\partial_{\tau}-\mu)^2-\nabla^2
\Bigr]\varphi(x)
=\frac{1}{2} 
\int_{p}
\Phi^{\dagger}(p)G_{0}^{-1}\Phi(p),\\
&\quad G_{0}^{-1}
:=
\begin{pmatrix}
-({\rm i}\omega_{n}+\mu)^2+\boldsymbol{p}^2 & 0 \\
0 & -({\rm i}\omega_{n}-\mu)^2+\boldsymbol{p}^2 \\
\end{pmatrix},
\end{split}
\end{equation}
where we employ the bilinear form for the later convenience.
Let us now take the homogeneous limit with $\varphi(x) \to \phi$.
The vertex function is then straightforwardly obtained as
\begin{equation}
[\Gamma_k^{(2)}(p,q)]_{ij}
= \delta^{(4)}(p-q)
    \Bigl[
        (G_0^{-1}(p))_{ij} + \delta_{ij} V'_k + \Phi_i \Phi_j^* V_k''
    \Bigr].
\end{equation}
Computing the inverse matrix $(\Gamma_{k}^{(2)}+R_{k})^{-1}$ with the Litim regulator~\eqref{eq:Litim}, we reduce the trace part of Eq.~\eqref{eq:Wfloweq} to the following form:
\begin{equation}
\begin{split}
{\rm Tr}\left[
\frac{\partial R_{k}}{\partial k}\frac{1}{\Gamma_{k}^{(2)}+R_{k}}
\right]
&= 4\beta V k \int_p \frac{\theta(k-|\boldsymbol{p}|) \,\mathrm{Re} X}{|X|^2-|Y|^2},
\label{eq:trace}
\end{split}
\end{equation}
where the integrand are presented with
\begin{equation}
\begin{split}
X &:=(\omega_{n_p}-{\rm i}\mu)^2+E_{k}^2,
\quad
Y:=\varphi^2 U_{k}''.
\end{split}
\end{equation}
In the flow equation~\eqref{eq:floweq}, the overall factor $\beta V$ in Eq.~\eqref{eq:trace} is canceled by the one in $\Gamma_k = \beta V U_k$.
Thanks to the Litim regulator, the spacial momentum integral yields $4\pi k^3/3$.
The Matsubara summation is carried out with the four poles $z = \pm E_\pm$, where $E_\pm$ are defined by Eq.~\eqref{eq:Es}.
After all of these calculations, we eventually obtain the flow equation~\eqref{eq:floweq}.
\vspace{10pt}
\section*{Acknowledgments}
K.~M. is supported by the Japan Society for the Promotion of Science (JSPS) KAKENHI under Grant No.~24K17052.
F.~T. acknowledges support from the Moritani Scholarship Foundation (No.~22051) and JST SPRING under Grant No.~JPMJSP2151.

\vspace{-5pt}

\bibliography{frg}


\end{document}